\documentclass[pre,twocolumn,preprintnumbers,amsmath,amssymb]{revtex4}

\usepackage{graphicx}
\usepackage{dcolumn}
\usepackage{bm}
\usepackage{parskip}
\usepackage{subfigure}
\usepackage{amssymb}
\usepackage{amsmath}
\usepackage{amssymb}
\usepackage{graphicx}

\def\be{\begin{eqnarray}}
\def\ee{\end{eqnarray}}
\def\be{\begin{equation}}
\def\ee{\end{equation}}
\begin{document}
\title{Multiparameter universality and directional nonuniversality of exact anisotropic critical correlation functions of the two-dimensional Ising universality class}

\author{Volker Dohm}

\affiliation{Institute for Theoretical Physics, RWTH Aachen
University, D-52056 Aachen, Germany}

\date {16 October 2019}

\begin{abstract}
We prove the validity of multiparameter universality for the exact critical bulk correlation functions of the anisotropic square-lattice and triangular-lattice Ising models on the basis of the exact scaling structure of the correlation function of the two-dimensional anisotropic scalar $\varphi^4$ model with four nonuniversal parameters. The correlation functions exhibit a directional nonuniversality due to principal axes whose orientation depends on microscopic details. We determine the exact anisotropy matrices governing the bulk and finite-size critical behavior of the $\varphi^4$ and Ising models. We also prove the validity of multiparameter universality for an exact critical bulk amplitude relation.

\end{abstract}
\maketitle
The concept of bulk universality classes plays a fundamental role in the theory of critical phenomena \cite{fish-1, priv, pelissetto}. They are characterized by the spatial dimension $d$ and the symmetry of the ordered state which we assume here to be $O(n)$ symmetric with an $n$-component order parameter. Within a given $(d,n)$ universality class, critical exponents
and bulk scaling functions are independent of microscopic details, such as the couplings of (short-range) interactions or the lattice structure. It was asserted that, once the universal quantities of a universality class are given, two-scale-factor universality \cite{priv, pelissetto,stau,hohenberg1976} implies that the asymptotic (small $t=(T-T_c)/T_c$ ) critical behavior of any particular system
of this universality class is known completely provided that only two nonuniversal amplitudes are specified. It has been shown \cite{cd2004, dohm2006, dohm2008,kastening-dohm}, however,  that it is necessary to distinguish subclasses of isotropic and weakly anisotropic systems within a universality class and that two-scale-factor universality is not valid for the subclass of weakly anisotropic systems. In the latter systems there exists no unique bulk correlation-length amplitude but rather $d$ independent nonuniversal amplitudes in the $d$ principal directions. This has a significant effect on the anisotropic bulk order-parameter correlation function $G({\bf x},t)$ but a clear classification of its universality properties was not developed \cite{dohm2008}.

Recently \cite{dohm2018} the notion of {\it multiparameter universality}, originally introduced for critical amplitude relations \cite{dohm2008}, was formulated for the scaling structure of $G({\bf x},t)$ within the anisotropic $\varphi^4$ theory where $G({\bf x},t)$ depends on up to $d(d+1)/2+1$ independent nonuniversal parameters in $d$ dimensions, i.e., up to four or seven parameters in two or three dimensions, respectively. It was hypothesized that multiparameter universality of $G({\bf x},t)$ is valid not only for "soft-spin" $\varphi^4$ models but also for all weakly anisotropic systems within a given universality class including fixed-length spin models such as Ising $(n=1)$, $XY$ $(n=2)$, and Heisenberg $(n=3)$ models. No general proof was given for this hypothesis, except for an  analytic verification for a special example within the two-dimensional anisotropic Ising model at $T=T_c$ \cite{Wu1966}.

A unique opportunity for a significant test of the validity of multiparameter universality is provided by an analysis of the exact results for the bulk correlation function of the anisotropic "square-lattice"
and "triangular-lattice" Ising models \cite{WuCoy,Vaidya1976}
in the asymptotic scaling region near $T_c$.
Such an analysis is made possible by deriving the exact scaling structure of $G({\bf x},t)$ of the general anisotropic two-dimensional scalar $\varphi^4$ lattice model which belongs to the same universality class as the $d=2$ Ising model \cite{Mehlig}. We introduce angular-dependent correlation lengths which permits us to determine the principal axes via an extremum criterion and to derive the exact anisotropy matrices. This leads to a proof of multiparameter universality for the Ising models  with three or four nonuniversal parameters for the square-lattice or triangular-lattice model, respectively.
However, the correlation functions exhibit a directional nonuniversality due to the principal axes whose orientation depends on microscopic details. This dependence is different for Ising and $\varphi^4$ models. Our results are expected to make an impact on scaling theories for $G({\bf x},t)$ of real anisotropic systems such as magnetic materials \cite{alpha}, superconductors \cite{schneider2004}, alloys \cite{onukiBook}, and solids with structural phase transitions \cite{bruce-1}, where angular-dependent correlation functions are measurable quantities. Multiparameter universality is relevant also for finite-size effects, e.g., the critical Casimir force \cite{dohm2018}.

It is necessary to first reformulate the standard scaling form of  $G_\pm^{\text iso}({\bf x},t)$ for isotropic systems. In the limit of
large $| {\bf x} |$ and large $\xi^{\text iso}_{\pm}(t)=\xi^{\text iso}_{0\pm}|t|^{-\nu}$ at fixed  $|{\bf x}| / \xi^{\text iso}_{\pm}\geq 0$ the scaling form for $n=1$ and $2\leq d<4$ reads  \cite{pri,pelissetto}
\begin{eqnarray}
\label{3c} &&G_\pm^{\text iso}({\bf x},t) = D^{\text iso}_1 | {\bf x} |^{- d + 2 -
\eta} \Phi_\pm (|{\bf x}| / \xi^{\text iso}_{\pm}), \;\;\;\;\;\;\;
\end{eqnarray}
with the universal scaling function $\Phi_\pm (y)$ above $(+)$ and below ($-$) $T_c$ and the nonuniversal amplitudes $D^{\text iso}_1,  \xi^{\text iso}_{0 \pm}$, where  $\xi^{\text iso}_{0 +}/\xi^{\text iso}_{0 -}$ is universal but $D^{\text iso}_1$ still contains a universal part. We employ the "true" (exponential) correlation lengths $\xi^{\text iso}_{\pm}$ which are defined by the exponential decay $\sim {\text exp}\;(-|{\bf x}| / \xi^{\text iso}_{\pm})$ of $ \Phi_\pm$ for large $|{\bf x}| / \xi^{\text iso}_{\pm}$ and which are universally related to the second-moment correlation lengths \cite{pelissetto}.
The exact sum rule \cite{dohm2008} $\chi_\pm^{\text iso}(t)=\int d^d{\bf x}\;G_\pm^{\text iso}({\bf x},t)$ yields the susceptibility
\begin{eqnarray}
\label{a1} \chi_\pm^{\text iso} (t) &=& D^{\text iso}_1 \; \big[\xi_\pm^{\text iso} (t)\big ]^{2 - \eta} \; \widetilde \Phi_\pm
=\Gamma^{\text iso}_\pm |t|^{-\gamma}\;
,\\
\label{a2} \widetilde \Phi_\pm &=& 2 \pi^{d/2} \Gamma (d/2)^{-1}
\int_0^\infty ds s^{1 - \eta} \Phi_\pm (s) \;,
\end{eqnarray}
with $2-\eta=\gamma/\nu$ and the universal quantities $\widetilde \Phi_+$, $\widetilde \Phi_-$, and $\Gamma^{\text iso}_+ / \Gamma^{\text iso}_- $. This implies $D^{\text iso}_1=\Gamma^{\text iso}_+\;(\xi^{\text iso}_{0+})^{-2+\eta}/\widetilde \Phi_+$, thus $G_\pm^{\text iso}$ can be uniquely divided into universal and nonuniversal parts
\begin{eqnarray}
\label{3cuni} &&G_\pm^{\text iso}({\bf x},t) = \Gamma^{\text iso}_+ (\xi_{0+}^{\text iso} )^{-2+ \eta} \;| {\bf x} |^{- d + 2 -
\eta} \Psi_\pm (|{\bf x}| / \xi^{\text iso}_{\pm}), \;\;\;\;\;\;\;\\
\label{3d}
&& \Psi_+ (y) =  \Phi_+ (y)/\widetilde \Phi_+, \;\Psi_- (y) =  \Phi_- (y)/\widetilde \Phi_+,
\end{eqnarray}
with two nonuniversal amplitudes $\Gamma^{\text iso}_+$ and $ \xi_{0+}^{\text iso}$ and the universal scaling function $\Psi_\pm (y)$.
At $T_c$ it is related to the universal constants \cite{dohm2008,priv,tarko} $\widetilde Q_3$ and $Q_3$  by
\begin{eqnarray}
\label{tildeQ}
\Psi_+ (0) = \Psi_- (0) =  \widetilde Q_3= \frac{2^{d-2+\eta}\Gamma[(d-2+\eta)/2]}{(4\pi)^{d/2})\Gamma[(2-\eta)/2]}\;Q_3.
\end{eqnarray}
For $d=2$, $\Gamma^{\text iso}_+$ and $\xi_{0+}^{\text iso}$ are related to the amplitude $B^{\text iso}$ of the order  parameter ${\cal M}^{\text iso}=B^{\text iso}|t|^{1/8}$  and to the specific-heat amplitude $A^{\text iso}$ through \cite{pelissetto}
\begin{eqnarray}
\label{amplreliso}
(B^{\text iso})^2 (\Gamma^{\text iso})^{-1}(\xi^{\text iso}_{0+})^2= Q_c
\end{eqnarray}
and $A^{\text iso}(\xi^{\text iso}_{0 +})^2=(R^+_\xi)^2$
where $Q_c$ and $R^+_\xi=(2\pi)^{-1/2}$ are universal constants according to Eqs. (2.50),(3.49), and (6.31) of \cite{priv}. We present the exact value of $Q_c$ in (\ref{QB}) below.
Our only assumption is the validity of two-scale-factor universality for isotropic systems which implies that $\Psi_\pm,Q_c,R^+_\xi$ are the same for isotropic Ising and $\varphi^4$ models with $\gamma=2-\eta=7/4,\nu=1$, and $\xi^{\text iso}_{0 +}/\xi^{\text iso}_{0 -}=2$.

We first consider the anisotropic scalar $\varphi^4$ model on $\widetilde N$ lattice points ${\bf x}_i \equiv (x_{i1}, x_{i2})$ of a square lattice with lattice spacing $\tilde a$ and finite-range interactions $K_{i,j}$. The Hamiltonian divided by $k_B T$ and the bulk correlation function are defined by \cite{dohm2008}
\begin{eqnarray}
\label{2a}&& H  =   \tilde a^2 \Bigg[\sum_{i=1}^{\widetilde N} \left(\frac{r_0}{2}
\varphi_i^2 + u_0 \varphi_i^4 \right)
 +\sum_{i, j=1}^{\widetilde N} \frac{K_{i,j}} {2} (\varphi_i -
\varphi_j)^2 \Bigg], \;\;\;\;\\
\label{2l} &&G_\pm({\bf x_i-x_j}\;, t) = \lim_{\widetilde N \to \infty}
[< \varphi_i \varphi_j>- {\cal M}^2],
\end{eqnarray}
where ${\cal M}^2 =  \lim_{|{\bf x_i-x_j}| \to \infty}< \varphi_i \varphi_j>$. The large-distance anisotropy is described by the anisotropy matrix
\begin{eqnarray}
\label{abc}
{\bf A}&=&( A_{\alpha \beta})
=
\left(\begin{array}{ccc}
 a & c \\
 c & b \\
\end{array}\right),\\
\label{2i} A_{\alpha \beta} &=&\lim_{\widetilde N\to \infty} {\widetilde N}^{-1} \sum^{\widetilde N}_{i,
j = 1} (x_{i \alpha} - x_{j \alpha}) (x_{i \beta} - x_{j \beta})
K_{i,j}\;\;\;\;
\end{eqnarray}
where weak anisotropy requires ${\det \bf  A}>0, a>0,  b>0$ which ensures unchanged critical exponents \cite{cd2004}.
It has been shown recently \cite{dohm2018} that  $G_\pm({\bf x}, t)$ has the asymptotic scaling form
\begin{eqnarray}
\label{Gneu} G_\pm({\bf x}, t) = \frac{\Gamma_+\;(\bar \xi_{0+})^{-7/4}}{ [{\bf x}\cdot {\bf \bar A}^{-1}{\bf x}]^{1/8}}\;
 \Psi_\pm \Big(\frac{[{\bf x}\cdot {\bf \bar A}^{-1}{\bf x}]^{1/2}}{\bar \xi_\pm(t)}\Big)\;\;\;
\end{eqnarray}
with ${\bf \bar A}={\bf  A}/({\det \bf  A})^{1/2}$ where  $\Psi_\pm$ is the same scaling function as that  in (\ref{3cuni}) for isotropic systems (${\bf \bar A}={\bf  1}$). We have obtained (\ref{Gneu}) from Eqs. (5.61) and (5.32) of  \cite{dohm2018} by employing the sum rule for the susceptibility of the anisotropic system $\chi_\pm(t)=\int d^2{\bf x}\;G_\pm({\bf x},t)=\Gamma_\pm |t|^{-7/4}$ which yields the nonuniversal constant $D_1=\Gamma_+\;(\bar \xi_{0+})^{-7/4}/\widetilde \Phi_+$ \cite{dohm2018}.
Here ${\bar \xi_\pm(t)}$ is the geometric mean
\begin{eqnarray}
\label{meancorl}{\bar \xi_\pm(t)}=\bar \xi_{0\pm}|t|^{-1},\;\; \bar \xi_{0\pm}=\big[\xi_{0\pm}^{(1)}\xi_{0\pm}^{(2)}\big]^{1/2}
\end{eqnarray}
of the principal correlation lengths $\xi_\pm^{(\alpha)}(t)= \xi^{(\alpha)}_{0\pm}|t|^{-1}$
where the principal axes are defined by the eigenvectors ${\bf e}^{(\alpha)}$ determined by ${\bf A e}^{(\alpha)}=\lambda_\alpha {\bf e}^{(\alpha)}, \alpha = 1,2$. The eigenvalues $\lambda_\alpha >0$ determine the amplitudes $ \xi_{0 \pm}^{(\alpha)} = \lambda_\alpha^{1/2} \xi_{0 \pm}'$ with $ \xi_{0 +}^{(\alpha)}/\xi_{0 -}^{(\alpha)} = \xi_{0 +}'/\xi_{0 -}'=2$  where $\xi_{0 \pm}'$ is the correlation length of the isotropic system obtained after a shear transformation that consists of a rotation and a rescaling in the ${\bf e}^{(\alpha)}$ directions \cite{cd2004, dohm2006, dohm2008,dohm2018}. The amplitudes $\xi_{0 +}^{(\alpha)}$ are independent of the amplitude $B$ of the order parameter ${\cal M}=B|t|^{1/8}$ of the anisotropic model. From the shear transformations  \cite{dohm2006,dohm2008,dohm2018} $(\xi'_{0+})^2=({\det \bf  A})^{-1/2}(\bar \xi_{0+})^2$, $(B')^2=({\det \bf  A})^{1/2}B^2, A'=({\det \bf  A})^{1/2} A$, and $\Gamma'_+=\Gamma_+$ we find the relations for the anisotropic system
\begin{eqnarray}
\label{amplrelaniso}
B^2 \;\Gamma_+^{-1}(\bar \xi_{0+})^2= Q_c
\end{eqnarray}
and $A(\bar \xi_{0+})^2=(R^+_\xi)^2$ where $Q_c$ and $R^+_\xi$ are the same as in the isotropic case. Thus the susceptibility amplitude $\Gamma_+$ is determined by {\it three} independent nonuniversal parameters $\xi_{0 +}^{(1)},\xi_{0 +}^{(2)},B$ whereas the specific-heat amplitude $A$ is determined by two parameters $ \xi_{0 +}^{(1)},\xi_{0 +}^{(2)}$, and $B^2$ can be expressed as $B^2= A\Gamma_+ Q_c/(R^+_\xi)^2$. The individual lengths $\xi^{(\alpha)}_{0+}$ cannot be determined from $A,B,$ and $\Gamma_+$.

Contours of constant correlations are ellipses determined by ${\bf x}\cdot{\bf\bar A}^{-1}{\bf x}= {\text const}$ whose excentricity and orientation are characterized by
\begin{eqnarray}
\label{corratio} q=(\lambda_1/\lambda_2)^{1/2}=\xi_{0\pm}^{(1)}/\xi_{0\pm}^{(2)}
\end{eqnarray}
and by the angle $\Omega$ determining
the principal axes, i.e.,
\begin{eqnarray}
 \label{2pp}
 {\bf e}^{(1)}& =& \left(\begin{array}{c}
  \cos\; \Omega \\
  \sin\; \Omega \\
\end{array}\right) ,\;\;\; {\bf e}^{(2)} = \left(\begin{array}{c}
  - \sin \;\Omega \\
  \cos\; \Omega  \\
\end{array}\right). \;\;\;
\end{eqnarray}
For $a \neq b$ we define
\begin{eqnarray}
\label{eigenvalue}
\lambda_{1} &=&\frac{a+b}{2}+ \frac{a-b}{2}\;w, \; \lambda_{2} =\frac{a+b}{2}- \frac{a-b}{2}\;w, \;\;\;\;\\
w&=&[1+4c^2/(a-b)^2]^{1/2}\geq 1,\\
\label{Omegaphi}
\tan \Omega&=&[b-a +(a-b) w]/(2c),
\end{eqnarray}
and for $a=b,c\neq 0$
\begin{eqnarray}
\label{eigenvaluex}
\lambda_{1} &=&a+c, \;\; \lambda_{2} =a-c, \;\;\Omega= \pi/4.
\end{eqnarray}
From ${\bf \bar A}={\bf U}^{-1}{\bf \bar{\mbox {\boldmath$\lambda$}}} {\bf U}$ with the rotation and rescaling matrices
${\bf U}=\left(\begin{array}{ccc}
  c_\Omega& s_\Omega \\
  -s_\Omega &  c_\Omega \\
\end{array}\right)$
and
${\bf \bar{\mbox {\boldmath$\lambda$}}}=
\left(\begin{array}{ccc}
 q& 0 \\
 0 & \;q^{-1} \\
\end{array}\right)$ we obtain
\begin{eqnarray}
\label{Aquer}
{\bf \bar A}(q,\Omega)=
\left(\begin{array}{ccc}
 q \;c_\Omega^2+q^{-1}s_\Omega^2 & \;\;\;(q-q^{-1})\;c_\Omega \;s_\Omega\\
(q-q^{-1})\; c_\Omega\; s_\Omega& q \;s_\Omega^2 +q^{-1}\;c_\Omega^2
\end{array}\right)
\end{eqnarray}
with the abbreviations $c_\Omega\equiv\cos\Omega,s_\Omega\equiv\sin\Omega$. Using polar coordinates ${\bf x}=(x_1, x_2)=(r\cos \theta, r \sin \theta)$ (Fig. 1)
we define the angular-dependent correlation length $\xi_\pm(t,\theta,q,\Omega)$ by
\begin{eqnarray}
\label{corrlength}
[{\bf x}\cdot ({\bf \bar A}^{-1}{\bf x})]^{1/2}/\bar \xi_\pm(t)=r/\xi_\pm(t,\theta,q,\Omega)
\end{eqnarray}
which yields the exact reformulation of (\ref{Gneu})
\begin{eqnarray}
\label{Gneuangular} G_\pm({\bf x}, t) &=& \frac{\Gamma_+\;(\bar \xi_{0+})^{-7/4}}{ [r f(\theta,q,\Omega)]^{1/4}}\;
 \Psi_\pm \Big(\frac{r}{\xi_\pm(t,\theta,q,\Omega)}\Big)\;
\end{eqnarray}
where the directional dependence is described by
\begin{eqnarray}
\label{corrlength}
\xi_\pm(t,\theta,q,\Omega)&=&\;\bar\xi_\pm(t)/f(\theta,q,\Omega),\\
 \label{angular}
f(\theta,q,\Omega)&=&  \big[ (q \;\sin^2 \Omega + q^{-1}\cos^2 \Omega)\;\cos^2\theta\nonumber \\&+&\;(q \;\cos^2 \Omega + q^{-1}\sin^2 \Omega)\;\sin^2\theta\nonumber \\
&+&\;(q^{-1} - q) \;\cos \Omega \;\sin \Omega \; \sin(2\theta) \big]^{1/2}\;\;\;\;\;\nonumber \\
&=&  \big[ q \;\sin^2(\theta-  \Omega) + q^{-1}\cos^2 (\theta-\Omega) \big]^{1/2}. \;\;\;\;\;\;\;
\end{eqnarray}
\vspace{0cm}
\begin{figure}
\includegraphics[clip,width=6.0cm]
{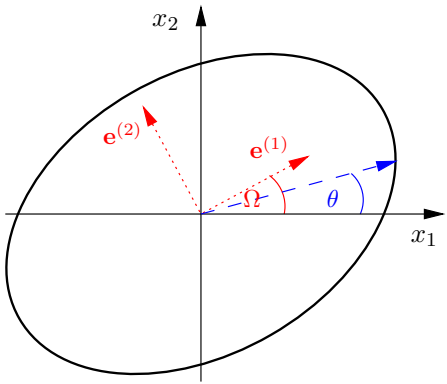}
 \vspace{1cm}
 \caption{(Color online) Elliptical contour of constant correlations of  $G_\pm({\bf x}, t)$, (\ref{Gneuangular})-(\ref{angular}). Dotted arrows ${\bf e}^{(1)}$ and ${\bf e}^{(2)}$: principal directions for  $a>b>0, c>0$ with $0<\Omega <\pi/4$. Dashed arrow: vector ${\bf x}$ in the direction $\theta$. }
\end{figure}
\vspace{0cm}
In the limit $c \to 0$ at fixed $a-b\neq 0$ a "rectangular" anisotropy is obtained with $ \lambda_1=a, \lambda_2=b,q=(a/b)^{1/2}=\xi_{0\pm}^{(1)}/\xi_{0\pm}^{(2)}$,  $\Omega=0$, and
\begin{eqnarray}
\label{angularrec}
f(\theta,q,0)\equiv  f_{\text rec}(\theta,q)=  \big(  q^{-1}\;\cos^2\theta +\;q \;\sin^2\theta \big)^{1/2}.
\end{eqnarray}
For $q\neq 1$ the requirement $\partial\xi_\pm(t,\theta,q,\Omega)/\partial \theta =0$ yields $\sin[2(\theta-\Omega)]=0$ implying that $\xi_\pm$ has extrema at $\theta^{(1)} =\Omega$ and $\theta^{(2)}=\Omega+\pi/2$ defining the two principal directions.

In contrast to $G^{\text iso}_\pm({\bf x}, t)$,  $G_\pm({\bf x}, t)$ depends on four independent nonuniversal parameters $\Gamma_+,\bar \xi_{0+},q,\Omega$ which violates two-scale-factor universality. Unlike $q$, the angle $\Omega(a,b,c)$ cannot be parameterized in terms of $\xi_{0\pm}^{(\alpha)}$ but depends on the lattice structure and the microscopic couplings $K_{i,j}$ through $a,b,c$. Thus ${\bf \bar A}$ depends not only on bulk correlation lengths through $q$ but also on other microscopic details through $\Omega$.
The parametrization of (\ref{corratio})-(\ref{angularrec}) is valid in the unrestricted range $0<q<\infty$ above, at, and below $T_c$ \cite{eigen}. The same matrix ${\bf \bar A}$ also enters the finite-size critical behavior \cite{dohm2018}.
These results derived for a $\varphi^4$ model on a square lattice remain valid more generally for a $\varphi^4$ model with couplings $K_{i,j}$ on two-dimensional Bravais lattices \cite{dohm2008}.

The hypothesis of multiparameter universality \cite{dohm2018} predicts that the critical correlation functions of all anisotropic Ising models with short-range interactions can be expressed in the same form as (\ref{Gneuangular})-(\ref{angular}) with the same
universal functions $\Psi_\pm$ and $f$ and the same critical exponents,
but with up to four different nonuniversal parameters.
We shall show that this is indeed valid for Ising models with the Hamiltonian  \cite{WuCoy,Vaidya1976,stephenson}
\begin{eqnarray}
\label{IsingH}
H^{\text Is} =
\sum_{j, k} [-E_1\sigma_{j,k} \sigma_{j,k+1}&-&E_2\sigma_{j,k} \sigma_{j+1,k}\nonumber\\ &-&E_3\sigma_{j,k} \sigma_{j+1,k+1}]\;\;\;
\end{eqnarray}
where $\sigma_{j,k}=\pm1$ are spin variables  on a square lattice (with the lattice spacing $\tilde a = 1$)  with horizontal, vertical, and diagonal couplings $E_1>0, E_2>0$, $E_3$ (Fig.2). The exact correlation function $<\sigma_{0,0}\; \sigma_{M,N}>_\pm$  at vanishing external field was calculated for $E_3=0$ in \cite{WuCoy} and for positive and negative $E_3\neq 0$ in \cite{Vaidya1976}, resulting in the scaling form
\begin{eqnarray}
\label{IsingCorr}
<\sigma_{0,0}\; \sigma_{M,N}>_\pm=R^{-1/4}{\cal F}_\pm(R/\xi_\pm^{\text Is},E_1,E_2,E_3)
\end{eqnarray}
with a nonuniversal scaling function ${\cal F}_\pm$, a distance $R(E_1,E_2,E_3)$,
and a correlation length  $\xi_\pm^{\text Is}=\xi^{\text Is}_{0\pm}(E_1,E_2,E_3)|t|^{-1}$ with $\xi_+^{\text Is}/\xi_-^{\text Is}=2$.
The exact amplitude $C_{0\pm}(E_1,E_2,E_3)$ of the susceptibility was also calculated.
So far the universality properties of (\ref{IsingCorr}) have not been analyzed in the literature  \cite{priv, pelissetto,Vaidya1976,WuCoy,CoyWu}, and the universal part of the function ${\cal F}_\pm$ has not been identified. In particular, the principal axes and principal correlation lengths of the triangular-lattice model ($E_3\neq 0$) are as yet unknown, and only a conjecture exists for the correlation lengths in the direction of the bonds \cite{Indekeu}. For comparison with (\ref{2l}) below $T_c$ we need to consider the {\it subtracted} correlation function
\begin{eqnarray}
\label{IsingCorrsub}
<\sigma_{0,0}\; \sigma_{M,N}>_\pm ^{\text sub}=<\sigma_{0,0}\; \sigma_{M,N}>_\pm \;-\; ({\cal M}^{\text Is})^2
\end{eqnarray}
where ${\cal M}^{\text Is}(E_1,E_2,E_3)$ is the
spontaneous
magnetization, with  ${\cal M}^{\text Is}=0$ for $T\geq T_c$. We shall analyze three cases.
\vspace{0cm}
\begin{figure}
\includegraphics[clip,width=6.0cm]
{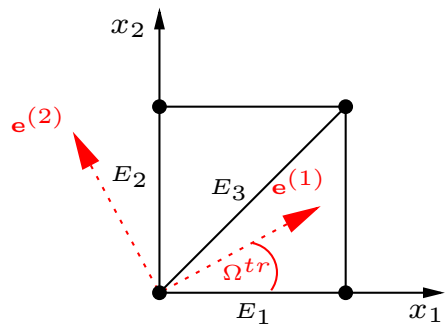}
 \vspace{0cm}
 \caption{(Color online)  Lattice points of the "triangular-lattice" Ising  model (\ref{IsingH}) on a square lattice with couplings $E_1,E_2,E_3$. Dotted arrows ${\bf e}^{(1)}$ and ${\bf e}^{(2)}$: principal directions for  $E_1>E_2>0, E_3>0$ with $0<\Omega^{\text tr} <\pi/4$. }
\end{figure}

We start from the isotropic case $E_1=E_2=E>0$, $E_3=0$,  where $R(E,E,0)=(M^2+N^2)^{1/2}\equiv R$,  and \cite{WuCoy}
\begin{eqnarray}
\label{corrlengthiso}
&&\xi^{\text Is}_{0+}(E,E,0)=(4\beta_cE)^{-1}=[2\ln(1+2^{1/2})]^{-1}\equiv\xi^{\text Is}_{0+},\\
\label{magnetisationiso}
&&{\cal M}^{\text Is}(E,E,0)=2^{5/16}[\ln(1+2^{1/2})]^{1/8}\;|t|^{1/8}\equiv B^{\text Is}|t|^{1/8},\;\;\;\;\;\;\;\\
\label{Isingsus}
&&C_{0+}(E,E,0)= 2^{19/8}\;\pi [\xi^{\text Is}_{0+}(E,E,0)]^{7/4} \;p_+\equiv C_{0+}^{\text Is}\;,\;\;\;\;\;\;\\
\label{pplus}
&&p_+=C_{0+}(E,E,0)/D = 0.1592846958...,
\end{eqnarray}
with $\beta_c=(k_BT_c)^{-1},  \sinh2\beta_cE=1$.
The constant $p_+$ is expressed analytically in terms of a Painlev\'e function of the third kind and its numerical value follows from Eq. (2.52S) of \cite{WuCoy}.
According to (\ref{3cuni})
we reformulate (\ref{IsingCorrsub}) as
\begin{eqnarray}
\label{Isingscal}
&&<\sigma_{0,0}\; \sigma_{M,N}>_\pm ^{\text sub}\;=\;\frac{C_{0+}^{\text Is}(\xi_{0+}^{\text Is})^{-7/4}}{R^{1/4}} \; \Psi_\pm(R/\xi_\pm^{\text Is}),\;\;\;\;\; \;\;\\
\label{PsiIsingplus}&&\Psi_+(y_+)=(2^{19/8}\;\pi \;p_+)^{-1}\widetilde F_+(y_+),\\
\label{PsiIsingminus}&&\Psi_-(y_-)=(2^{19/8}\;\pi \;p_+)^{-1} \nonumber\\
&&\;\;\;\;\;\;\;\;\;\;\;\;\;\;\;\times \;\big[\widetilde F_-(y_-/2)\;-\;2^{3/8}(y_-/2)^{1/4}\;],
\end{eqnarray}
with  $y_\pm=R/\xi^{\text Is}_{\pm}$,  $\xi_\pm^{\text Is}\equiv \xi^{\text Is}_{0\pm}(E,E,0)|t|^{-1}$ where the functions $\widetilde F_+(y_+)$ and $\widetilde F_-(y_-/2)$ are given by the right-hand side of Eq. (2.39) of \cite{WuCoy} with $\sinh2\beta_cE_1+\sinh2\beta_cE_2$ replaced by $2$ and with the argument $\theta$ replaced by $y_+/2$ or $y_-/4$, respectively. The unsubtracted correlation function \cite{WuCoy} $<\sigma_{0,0}\; \sigma_{M,N}>_\pm$  is easily obtained by dropping the last term in (\ref{PsiIsingminus}) which comes from $-({\cal M}^{\text Is})^2$ in (\ref{IsingCorrsub}). According to two-scale-factor universality the functions $\Psi_\pm$ identify the exact universal scaling functions above and below $T_c$ of all systems in the subclass of {\it isotropic} systems in the $(d=2,n=1)$ universality class. In particular,  $\Psi_+(0)=\Psi_-(0)=\widetilde Q_3$ is a universal amplitude.
Its exact value is $\widetilde Q_3=0.270969...$
where we have used Eq. (5.10S) of \cite{WuCoy}. This implies $Q_3=0.414131...$.

Now we turn to the case of a "rectangular" anisotropy $E_1\neq E_2$, $E_3=0$ where the condition of criticality is $S_1S_2=1$ with $S_\alpha= \sinh 2\beta^{\text rec}_cE_\alpha, \alpha=1,2$ \cite{WuCoy}. Using $x_1=N=r \cos\theta, x_2= M  =r\sin\theta$ we derive from Eqs. (2.6), (2.8), (2.10), and (2.44) of  \cite{WuCoy}
\begin{eqnarray}
 \label{qIsing}
&& R(E_1,E_2,0)=\big[q^{\text rec}M^2+(q^{\text rec})^{-1}N^2\big]^{1/2}\\
&&\;\;\;\;\;\;\;\;\;\;\;\;\;\;\;\;\;\;\;\;=r f_{\text rec}(\theta,q^{\text rec}),\\
&&q^{\text rec}=  \xi_{0\pm}^{(1){\text rec}}/\xi_{0\pm}^{(2){\text rec}}= (S_1/S_2)^{1/2},\\
\label{corrlengthrecplus}
&&\xi_{0+}^{(\alpha){\text rec}}= S_\alpha^{1/2}\;\big[2\beta^{\text rec}_c E_1/S_1^{1/2} + 2\beta^{\text rec}_c E_2/S_2^{1/2}\big]^{-1}\;\;\;\;\\
\label{corrlengthrecminus}
&&\;\;\;\;\;\;\;\;\;=2\;\xi_{0-}^{(\alpha){\text rec}},\\
\label{magnrec}
&&{\cal M}^{\text Is}(E_1,E_2,0)=\big[q^{\text  rec}+(q^{\text  rec})^{-1}\big]^{1/16}\big[\bar \xi_-^{{\text \;rec}}(t)\big]^{-1/8}\;\;\;\;\;\;\;\;\\
&&\;\;\;\;\;\;\;\;\;\;\;\;\;\;\;\;\;\;\;\;\;\equiv B^{\text rec}\;|t|^{1/8},\\
\label{meancorrrec}
&&\bar \xi_\pm^{{\text \;rec}}(t)=\bar \xi_{0\pm}^{{\text \;rec}}|t|^{-1},\;\; \bar \xi_{0\pm}^{{\text \;rec}}=\big[\xi_{0\pm}^{(1){\text rec}}\xi_{0\pm}^{(2){\text rec}}\big]^{1/2},
\end{eqnarray}
where $\xi_{0\pm}^{(1){\text rec}}$ and $\xi_{0\pm}^{(2){\text rec}}$ are the correlation-length amplitudes in the  principal directions $(1,0)$ and $(0,1)$, respectively, corresponding to $\Omega^{\text rec}=0$ and where $f_{\text rec}(\theta,q^{\text rec})$ is the same function as defined in (\ref{angularrec}), with $q$ replaced by $q^{\text rec}$. From  Eq. (2.8) of  \cite{WuCoy} we derive
\begin{eqnarray}
\label{meancorrlengthrec}
\xi_{0+}^{{\text Is}}(E_1,E_2,0)&=& [2\beta^{\text rec}_c(E_1C_1+E_2C_2)]^{-1}\nonumber\\\times(S_1+S_2)^{1/2} &=&\bar \xi_{0+}^{{\text \;rec}}\\
&=&2\;\xi^{\text Is}_{0-}(E_1,E_2,0) =2\bar \xi_{0-}^{{\text \;rec}}
\end{eqnarray}
where $C_\alpha= \coth 2\beta^{\text rec}_cE_\alpha$. This identifies "the correlation length $\xi$" in \cite{WuCoy} as  $\bar \xi_+^{{\text \;rec}}(t)$ above $T^{{\text \;rec}}_c$ and $2\bar \xi_-^{{\text \;rec}}(t)$ below $T^{{\text \;rec}}_c$ as follows from Eq. (2.31) of \cite{WuCoy}. From Eqs. (2.46a) and (2.48) of  \cite{WuCoy} we obtain the exact amplitude $C^{\text rec}_{0+}\equiv C_{0+}(E_1,E_2,0)$  in the form
\begin{eqnarray}
\label{susrectang}
C^{\text rec}_{0+}= 2^{9/4}\pi p_+(\bar \xi_{0+}^{{\text \;rec}})^{7/4}\big[q^{\text  rec}+(q^{\text  rec})^{-1}\big]^{1/8},
\end{eqnarray}
and from Eq. (2.39) of  \cite{WuCoy} we obtain
\begin{eqnarray}
\label{amplrel}
{\cal F}_\pm(y,E_1,E_2,0)=\big\{\big[q^{\text rec}+(q^{\text rec})^{-1}\big]/2\big\}^{1/8}{\cal F}_\pm(y,E,E,0).\;\;
\end{eqnarray}
Together with (\ref{PsiIsingplus}) and  (\ref{PsiIsingminus}) this leads to the exact reformulation of the asymptotic result of \cite{WuCoy}
\begin{eqnarray}
\label{corIsingrec}
&&<\sigma_{0,0}\; \sigma_{M,N}>_\pm^{\text rec,sub}\nonumber\\
 &&= \frac{C^{\text rec}_{0+}\;(\bar \xi_{0+}^{{\text \;rec}})^{-7/4}}{ [{\bf x}\cdot ({\bf \bar A}^{\text rec})^{-1}{\bf x}]^{1/8}}\;
 \Psi_\pm \Big(\frac{[{\bf x}\cdot ({\bf \bar A}^{\text rec})^{-1}{\bf x}]^{1/2}}{\bar \xi^{\text\; rec}_\pm(t)}\Big)\;\;\;\\
 \label{corIsingrecf}
&&=\frac{C^{\text rec}_{0+}\;(\bar \xi_{0+}^{{\text \;rec}})^{-7/4}}{ [r f(\theta,q^{\text rec},0)]^{1/4}}\;
\Psi_\pm \Big(\frac{r}{\xi^{\text rec}_\pm(t,\theta)}\Big),\;\;\;\;\;\;\\
\label{corlengthrec}
&& \xi^{\text rec}_\pm(t,\theta)=
 \bar \xi_{\pm}^{{\text \;rec}}(t)/f(\theta,q^{\text rec},0),
\end{eqnarray}
with the angular-dependent correlation lengths $\xi^{\text rec}_\pm(t,\theta)$ and with ${\bf \bar A}^{\text rec}(q^{\text rec})={\bf \bar A}(q^{\text rec},0)$, in exact agreement with (\ref{Gneu}) and (\ref{Aquer})-(\ref{angularrec}) for $\Omega^{\text rec}= 0$, thus confirming multiparameter universality above, at, and below $T^{\text rec}_c$ with three nonuniversal parameters $C^{\text rec}_{0+},\bar \xi_{0+}^{{\text \;rec}},q^{\text rec}$. This is valid for both $E_1\geq E_2$ and $E_1\leq E_2$ in the unrestricted range $0<q^{\text rec} < \infty$. For $E_1=E_2$ the isotropic results are recovered.

We proceed to the case of a "triangular" anisotropy \cite{Vaidya1976,stephenson} with $E_3\neq 0$ and the condition of criticality $\hat S_1 \hat S_2+\hat S_2 \hat S_3+\hat S_3 \hat S_1=1$ with $\hat S_\alpha= \sinh 2\beta^{\text tr}_c E_\alpha, \alpha=1,2,3$  \cite{Berker}. We first determine the angle $\Omega^{\text tr}$ describing the orientation of the principal axes. For $T\approx T^{\text tr}_c$ the angular dependence of the distance $R(E_1,E_2,E_3)\equiv R_{\text tr}(\theta)=r f_{\text tr}(\theta)$ of Eq. (11) of \cite{Vaidya1976} is given by
\begin{eqnarray}
\label{ftr}
f_{\text tr}(\theta)&=&\big[(\hat S_1+ \hat S_3 )\sin^2\theta+(\hat S_2+ \hat S_3 )\cos^2\theta\nonumber\\&-&\hat S_3\sin2\theta\big]^{1/2}.
\end{eqnarray}
We define the angular-dependent correlation length  $\xi^{\text tr}_{\pm}(t,\theta)$ by rewriting the scaled variable of Eq. (10) of \cite{Vaidya1976} for $T\approx T^{\text tr}_c$ as
\begin{eqnarray}
\label{scalargtr}
R^{\text tr}(\theta)/\bar \xi_{\pm}^{{\text \;tr}}(t)&=&r/\xi^{\text tr}_\pm(t,\theta),\\
\label{angularcor}
\xi^{\text tr}_\pm(t,\theta)&=&\bar \xi_{\pm}^{{\text \;tr}}(t)/f_{\text tr}(\theta).
\end{eqnarray}
The requirement $\partial\xi^{\text tr}_\pm(t,\theta)/\partial \theta =0$ yields $\theta=\Omega^{\text tr}$  with
\begin{eqnarray}
\label{omega1}
&&\tan 2\Omega^{\text tr}=\frac{2\hat S_3}{\hat S_1- \hat S_2}=\frac{2(1-\hat S_1\hat S_2)}{\hat S_1^2-\hat S_2^2}\;\; \mbox{for} \;\; E_1\neq E_2,\;\;\;\;\;\;\;
\\
\label{omega2}
&&\Omega^{\text tr}=\pi/4\;\;\;\;\;  \mbox{for} \;\; \;E_1=E_2,
\end{eqnarray}
implying that $\xi^{\text tr}_\pm(t,\theta)$ has extrema at $\theta^{(1)} =\Omega^{\text tr}$ and $\theta^{(2)}=\Omega^{\text tr}+\pi/2$  defining the two principal directions.
Clearly $\Omega^{\text tr}(E_1,E_2,E_3)$ is a nonuniversal quantity that depends on microscopic details and differs from $\Omega$, (\ref{Omegaphi}), of the $\varphi^4$ model even if the $\varphi^4$ and Ising models have the same couplings on the same lattices. This is due to the nonuniversal difference between a fixed-length spin model and a soft-spin model.
From (\ref{ftr})-(\ref{omega2}) we determine the ratio $q^{\text tr}=  \xi_{0\pm}^{(1){\text tr}}/\xi_{0\pm}^{(2){\text tr}}$ of the amplitudes of the principal correlation lengths
\begin{eqnarray}
\label{ratiocortr}
&&q^{\text tr}(E_1,E_2,E_3)=\frac{\xi^{\text tr}_\pm(t,\Omega^{\text tr})}{\xi^{\text tr}_\pm(t,\Omega^{\text tr}+\pi/2)}=\frac{f_{\text tr}(\Omega^{\text tr}+\pi/2)}{f_{\text tr}(\Omega^{\text tr})}\nonumber\\
\label{ratiocorSi}
&&
=\frac{2+\hat S_1^2+\hat S_2^2\pm \big[(\hat S_1^2+\hat S_2^2)^2+4(1-2\hat S_1\hat S_2)\big]^{1/2}}{2(\hat S_1+\hat S_2)}\;\;\;\;\;\;\;\;\;
\end{eqnarray}
where now the $\pm$ sign in front of the square root term means $E_1 > E_2$ (+) and $E_1 < E_2 $ $(-)$, respectively, and
\begin{eqnarray}
\label{ratiocortrgleich}
q^{\text tr}(E,E,E_3)=\frac{f_{\text tr}(3\pi/4)}{f_{\text tr}(\pi/4)}= \frac{1}{\sinh 2\beta^{\text tr}_c E}
\end{eqnarray}
for $E_1 =E_2=E>0$. From (\ref{omega1})-(\ref{ratiocortrgleich}) we derive
\begin{eqnarray}
\label{angularfunc1}
&&q^{\text tr}\cos^2\Omega^{\text tr}+{q^{\text tr}}^{-1}\sin^2\Omega^{\text tr}=\hat S_1+\hat S_3,\\
\label{angularfunc2}
&&q^{\text tr}\sin^2\Omega^{\text tr}+{q^{\text tr}}^{-1}\cos^2\Omega^{\text tr}=\hat S_2+\hat S_3,\\
\label{angularfunc3}
&&({q^{\text tr}}^{-1}-q^{\text tr})\cos\Omega^{\text tr}\sin\Omega^{\text tr}=-\hat S_3,
\end{eqnarray}
for both $E_1 \geq E_2$  and $E_1 \leq E_2 $. Together with (\ref{ftr}) these equations prove the validity of the identification
$f_{\text tr}(\theta)=f(\theta,q^{\text tr},\Omega^{\text tr})$
in the unrestricted range \cite{smalllarge} $0<q^{\text tr} <\infty$ above, at, and below $T_c^{\text tr}$ where $f$ is indeed the same function (\ref{angular}) as derived within the $\varphi^4$ theory.
This completes the determination of the angular dependence of the anisotropy matrix
${\bf \bar A}^{\text tr}\equiv{\bf \bar A}(q^{\text tr},\Omega^{\text tr})$
for the triangular-lattice Ising model (\ref{IsingH}) where ${\bf \bar A}$ is the same matrix as in (\ref{Aquer}) for the $\varphi^4$ model,
in exact agreement with multiparameter universality.
Our results for rectangular anisotropy are recovered from (\ref{ftr})-(\ref{angularfunc3}) in the limit $E_3 \to 0$.

We mention two earlier conjectures.  (i) From (\ref{ftr}) and  (\ref{angularcor}) we derive
\begin{eqnarray}
\label{corrratios}
\kappa_1&=&\frac{\xi_{0\pm}^{({\text diag}){\text tr}}}{2^{1/2}\xi_{0\pm}^{(hor){\text tr}}}=\frac{f_{\text tr}(0)}{2^{1/2}f_{\text tr}(\pi/4)}=\frac{\cosh 2\beta^{\text tr}_c E_2}{\hat S_1+ \hat S_2},\;\;\;\\
\kappa_2&=&\frac{\xi_{0\pm}^{({\text diag}){\text tr}}}{2^{1/2}\xi_{0\pm}^{(vert){\text tr}}}=\frac{f_{\text tr}(\pi/2)}{2^{1/2}f_{\text tr}(\pi/4)}=\frac{\cosh 2\beta^{\text tr}_c E_1}{\hat S_1+\hat S_2},\;\;\;\;\;\;
\end{eqnarray}
where $\xi_{0\pm}^{({\text diag}){\text tr}}$
$,\xi_{0\pm}^{({\text hor}){\text tr}}$, and $\xi_{0\pm}^{({\text vert}){\text tr}}$
denote the correlation lengths in the $(1,1)$
$,(1,0)$, and $(0,1)$ directions
and the factor $2^{1/2}$ accounts for the diagonal lattice spacing. This confirms the conjecture in Eq. (2.6) of \cite{Indekeu}.
(ii) The ratio (\ref{ratiocortrgleich}) used in Sec. V. C of \cite{dohm2018} was based on the conjecture in Eq. (A22) of \cite{Indekeu} and is derived here directly from the exact result (\ref{ftr}).

In the remaining analysis of the triangular case we confine ourselves to $E_1=E_2=E_3=E>0$
where
\begin{eqnarray}
\label{QqEE}
&&\Omega^{\text tr}(E,E,E)=\pi/4,\;\\
\label{qtrEE}
&&q^{\text tr}(E,E,E)=1/\sinh2\beta^{\text tr}_c E=3^{1/2},\\
&&f_{\text tr}(\theta)=f(\theta,3^{1/2},\pi/4)= 3^{-1/4}(2-\sin2\theta)^{1/2}.\;\;
\end{eqnarray}
By expanding Eqs. (2) and  (10) of \cite{Vaidya1976} around $T^{\text tr}_c$ to leading order in $|t|=|T-T^{\text tr}_c|/T^{\text tr}_c$ we determine the magnetization, the mean correlation lengths $\bar \xi_{\pm}^{{\text \;tr}}(t)=\bar \xi_{0\pm}^{{\text \;tr}}|t|^{-1}$, and the principal correlation lengths $\xi_{0+}^{(\alpha){\text tr}}$ as
\begin{eqnarray}
\label{magnetizationtr}
&&{\cal M}^{\text Is}(E,E,E)=(4\ln 3)^{1/8}\;|t|^{1/8}\equiv B^{{\text tr}}\;|t|^{1/8},\\
\label{xibartr}
&&\bar \xi_{0+}^{{\text \;tr}}=[\xi_{0+}^{(1){\text tr}}\xi_{0+}^{(2){\text tr}}]^{1/2}=3^{-3/4}2^{1/2}/\ln 3\;\;\;\;\;\;\\
&&=3^{-1/4}\xi_{0+}^{(1){\text tr}}=3^{1/4}\xi_{0+}^{(2){\text tr}}=2\bar \xi_{0-}^{{\text \;tr}}.\;\;\;\;\;\;\;\;\;\;\;\;
\end{eqnarray}
From Eqs. (12) and (14)-(16) of \cite{Vaidya1976} we obtain
\begin{eqnarray}
\label{susctr}
C_{0+}(E,E,E)&=& 2^{21/8}3^{-3/16}\pi p_+(\bar \xi_{0+}^{{\text \;tr}})^{7/4}\equiv C^{\text tr}_{0+},\;\;\;\\
\label{amplreltriang}
{\cal F}_\pm(y,E,E,E)&=&2^{1/4}3^{-3/16} {\cal F}_\pm(y,E,E,0).
\end{eqnarray}
Together with (\ref{PsiIsingplus}) and (\ref{PsiIsingminus}) this leads to the exact reformulation of the asymptotic result of \cite{Vaidya1976}
\begin{eqnarray}
\label{corIsingtr}
&&<\sigma_{0,0}\; \sigma_{M,N}>_\pm^{\text tr,sub}\nonumber\\
 &&= \frac{C^{\text tr}_{0+}\;(\bar \xi_{0+}^{{\text \;tr}})^{-7/4}}{ [{\bf x}\cdot ({\bf \bar A}^{\text tr})^{-1}{\bf x}]^{1/8}}\;
 \Psi_\pm \Big(\frac{[{\bf x}\cdot ({\bf \bar A}^{\text tr})^{-1}{\bf x}]^{1/2}}{\bar \xi^{\text\; tr}_\pm(t)}\Big)\;\;\;\\
 \label{corIsingtrf}
&&=\frac{C^{\text tr}_{0+}\;(\bar \xi_{0+}^{{\text \;tr}})^{-7/4}}{ [r f(\theta,q^{\text tr},\Omega^{\text tr})]^{1/4}}\;
\Psi_\pm \Big(\frac{r}{\xi^{\text tr}_\pm(t,\theta,\Omega^{\text tr})}\Big),\;\;\;\;\;\;\\
\label{corlengthIsingtr}
&& \xi^{\text tr}_\pm(t,\theta,\Omega^{\text tr})=
 \bar \xi_{\pm}^{{\text \;tr}}(t)/f(\theta,q^{\text tr},\Omega^{\text tr}),
\end{eqnarray}
with the same universal functions $\Psi_+$, $\Psi_-$, and $f$ as in (\ref{Gneuangular})-(\ref{angular}), (\ref{PsiIsingplus}), and (\ref{PsiIsingminus}) and the same matrix ${\bf \bar A}^{\text tr}\equiv{\bf \bar A}(q^{\text tr},\Omega^{\text tr})$ as in (\ref{Aquer}), with the four nonuniversal parameters $C^{\text tr}_{0+},\bar \xi_{0+}^{{\text \;tr}},q^{\text tr},\Omega^{\text tr}$ given in (\ref{susctr}), (\ref{xibartr}), (\ref{qtrEE}), (\ref{QqEE}), respectively, thus proving the validity of multiparameter universality for the triangular-lattice Ising model
above, at, and below $T_c^{\text tr}$, and disproving two-scale-factor universality. Our hypothesis of multiparameter universality predicts  the structure of (\ref{corIsingtr})-(\ref{corlengthIsingtr}) to be valid also in the general case $E_1\neq E_2, E_3\neq 0$.

In order to complete our analysis we show that the universal amplitude relations (\ref{amplreliso}) and (\ref{amplrelaniso}) derived for the $\varphi^4$ model remain valid also for the Ising model. We first employ (\ref{corrlengthiso})-(\ref{pplus}) for the {\it isotropic} Ising model to derive
\begin{eqnarray}
\label{QBiso}
(B^{\text Is})^2 (C^{\text Is}_{0+})^{-1}(\xi_{0+}^{\text \;Is})^2=(4\pi p_+)^{-1},
\end{eqnarray}
in structural agreement with (\ref{amplreliso}). Thus our analysis identifies the exact universal constant $Q_c$ for $d=2,n=1$ as
\begin{eqnarray}
\label{QB}
Q_c= (4\pi p_+)^{-1}=0.499592701...  .
\end{eqnarray}
This can be confirmed by means of a different derivation from Eqs. (6.29) and (6.31) of \cite{priv}  which determines $Q_c= (R^+_\xi)^2/R_C$. From the rectangular and triangular results (\ref{magnrec})-(\ref{susrectang}) and (\ref{magnetizationtr})-(\ref{susctr}), respectively, we derive
\begin{eqnarray}
\label{amplrelrec}
(B^{\text rec})^2 (C^{\text rec}_{0+})^{-1}(\bar \xi_{0+}^{{\text \;rec}})^2&=&  (4\pi p_+)^{-1}=Q_c,\\
\label{amplreltr}
(B^{\text tr})^2 \;(C^{\text tr}_{0+})^{-1}(\bar \xi_{0+}^{\text \;tr})^2&=& (4\pi p_+)^{-1}= Q_c,
\end{eqnarray}
which agrees with (\ref{amplrelaniso}) for the anisotropic  $\varphi^4$ model. Thus both the anisotropic Ising and $\varphi^4$ models have universal amplitude relations with the same universal constant $Q_c$ as for the isotropic models, in agreement with the hypothesis of multiparameter universality. In the anisotropic cases (\ref{amplrelrec}) and
(\ref{amplreltr}) three independent nonuniversal parameters are involved for the same reasons as given in the context of (\ref{amplrelaniso}).

Multiparameter universality for other critical bulk amplitude relations within $\varphi^4$ theory in $d$ dimensions follows from Sec. III of \cite{dohm2008}, e.g., Eqs. (3.32)-(3.36). In particular, multiparameter universality is predicted, for general $n$, for anisotropic systems at $T_c$ in the presence of an ordering field $h$  with the amplitude $\Gamma_c$ of the susceptibility and the principal correlation lengths $\xi_c^{(\alpha)}$ according to Eq. (3.35) of \cite{dohm2008} for each $\alpha$, with a universal constant $Q_2(d,n)$ that is the same as for the corresponding relation \cite{tarko} of isotropic systems at $T_c$ in the same $(d,n)$ universality class. A verification of such relations within anisotropic fixed-length spin models would be interesting.

To summarize, we  have determined the exact anisotropy matrix ${\bf \bar A}$ for  anisotropic $\varphi^4$ and Ising models \cite{WuCoy,Vaidya1976} and have confirmed the validity of multiparameter universality for the exact bulk order-parameter correlation functions of these models above, at, and below $T_c$, thereby answering the longstanding question \cite{WuCoy} as to the universality properties of the Ising models. It is reassuring that the leading scaling part of the detailed expressions for $<\sigma_{0,0}\; \sigma_{M,N}>_\pm$  presented in \cite{WuCoy,Vaidya1976} can be condensed into the same compact universal forms (\ref{corIsingrecf}) and (\ref{corIsingtrf}) as the exact result (\ref{Gneuangular}) for the anisotropic  $\varphi^4$ model, with three universal functions $\Psi_+$, $\Psi_-$, and $f$. We have also found agreement with multiparameter universality for the exact critical bulk amplitude relations (\ref{amplrelaniso}), (\ref{amplrelrec}), and (\ref{amplreltr}) with three independent nonuniversal parameters.
These results support the validity of multiparameter universality for the large class of weakly anisotropic systems within the $(d,n)$ universality classes which is of relevance for studying the correlation functions in real anisotropic systems \cite{alpha,schneider2004,onukiBook,bruce-1}. The significance of multiparameter universality for finite-size effects, e.g.,  on the critical Casimir force and the specific heat, has been pointed out in \cite{dohm2018}. In all cases the universal critical exponents are not changed by weak anisotropy \cite{cd2004,dohm2008}, unlike the case of strong anisotropy \cite{tonchev}.
Nonuniversality enters $<\sigma_{0,0}\; \sigma_{M,N}>_\pm$ through the anisotropy matrix ${\bf \bar A}$, the mean correlation length, and the susceptibility  amplitude in the prefactor. ${\bf \bar A}$ is temperature-independent and is applicable above, at, and below $T_c$ in bulk and confined systems \cite{dohm2018}. As an appropriate parametrization of ${\bf \bar A}$  we have employed the ratio $q$ of the principal correlation lengths  and the angle $\Omega$ determining the principal directions. Both parameters are nonuniversal microscopic quantities. While for $\varphi^4$ models $\Omega$ is known explicitly according to (\ref{2i}) and (\ref{Omegaphi}),
this is not generally the case for Ising models. We agree with the assertion \cite{night1983}  that, apart from the Ising models \cite{WuCoy,Vaidya1976} analyzed in this paper, the principal directions "generically depend in an unknown way on the anisotropic interactions." Since the principal directions enter the angular dependence of correlation functions in a crucial way the unknown dependence  of $\Omega$ on microscopic details introduces a significant nonuniversality into the correlation functions of weakly anisotropic systems, in contrast to isotropic systems of the same universality class. This underscores the necessity of distinguishing subclasses of isotropic and anisotropic systems within a given  $(d,n)$ universality class. The latter are less universal than the former and require significantly more nonuniversal input in order to achieve quantitative predictions. This statement applies also to finite-size effects in anisotropic systems where up to  $d(d+1)/2+1$ nonuniversal parameters enter the finite-size scaling form of the free energy density \cite{dohm2008,dohm2018}. This sheds new light on the general belief that the critical behavior of systems with short-range interactions is largely independent of microscopic details \cite{fish-1,priv,pelissetto}.

\end{document}